\documentclass[aps,twocolumn,prd,superscriptaddress,preprintnumbers,showpacs]{revtex4-1}
\usepackage{graphicx}
\usepackage{bm}
\usepackage{amsmath}
\usepackage{amssymb}
\usepackage{amsthm}

\newcommand{\be}{\begin{equation}}
\newcommand{\ee}{\end{equation}}
\newcommand{\bea}{\begin{eqnarray}}
\newcommand{\eea}{\end{eqnarray}}
\newcommand{\bdm}{\begin{displaymath}}
\newcommand{\edm}{\end{displaymath}}
\newcommand{\beas}{\begin{eqnarray*}}
\newcommand{\eeas}{\end{eqnarray*}}

\begin{document}
\title{Further consistency tests of the stability of fundamental couplings}

\author{M. C. Ferreira}
\email[]{up080302013@alunos.fc.up.pt}
\affiliation{Centro de Astrof\'{\i}sica da Universidade do Porto, Rua das Estrelas, 4150-762 Porto, Portugal}
\author{C. J. A. P. Martins}
\email[]{Carlos.Martins@astro.up.pt}
\affiliation{Centro de Astrof\'{\i}sica da Universidade do Porto, Rua das Estrelas, 4150-762 Porto, Portugal}
\affiliation{Instituto de Astrof\'{\i}sica e Ci\^encias do Espa\c co, CAUP, Rua das Estrelas, 4150-762 Porto, Portugal}
\date{25 April 2015}

\begin{abstract}
In a recent publication [Ferreira {\it et al.}, Phys. Rev. D89 (2014) 083011] we tested the consistency of current astrophysical tests of the stability of the fine-structure constant $\alpha$ and the proton-to-electron mass ratio $\mu=m_p/m_e$ (mostly obtained in the optical/ultraviolet) with combined measurements of $\alpha$, $\mu$ and the proton gyromagnetic ratio $g_p$ (mostly in the radio band). Given the significant observational progress made in the past year, we now revisit and update this analysis. We find that apparent inconsistencies, at about the two-sigma level, persist and are in some cases enhanced, especially for matter era measurements (corresponding to redshifts $z>1$). Although hidden systematics may be the more plausible explanation, we briefly highlight the importance of clarifying this issue, which is within the reach of state-of-the art observational facilities such as ALMA and ESPRESSO.
\end{abstract}

\maketitle

\section{\label{intro}Introduction} 

Tests of the stability of nature's fundamental couplings are becoming an increasingly important probe of physics beyond the standard model \cite{RSoc,Uzan}. Very tight constraints stem from local laboratory tests using atomic clocks \cite{Rosenband}, while astrophysical measurements allow a large lever arm which can probe the dynamics of the new degrees of freedom responsible for such putative variations, and potentially shed light on the dark energy enigma \cite{Reconst}. There have been recent indications of possible variations \cite{Dipole}, which a dedicated Large Program at ESO's Very Large Telescope is aiming to test \cite{LP1,LP2,LP3}.

Direct astrophysical measurements of the fine-structure constant $\alpha$ and the proton-to-electron mass ratio $\mu=m_p/m_e$ are, in most cases, carried out the optical/ultraviolet (there are a few exceptions to this for the $\mu$ case), and up to redshifts now exceeding $z=4$. On the other hand, in the radio band, and typically at lower redshifts, one can measure various combinations of $\alpha$, $\mu$ and the proton gyromagnetic ratio $g_p$. In a recent work \cite{Frigola}
we carried out a joint statistical analysis of all existing data, in the context of a broad class of unification scenarios \cite{Coc}, and highlighted some apparent inconsistencies which could be an indication that systematics may be affecting some of the data.

Given the significant observational progress made in the past fifteen months, both in measurements of $\alpha$ \cite{LP3,Songaila} and in those of $\mu$ \cite{Albornoz,Bagdonaite4,BagdonaiteNew2}, in this work we will update our previous analysis. We start in Sect. II by briefly summarizing the theoretical assumptions underlying the class of unification scenarios we will assume, and then provide an up-to-date list of available measurements in Sect. III. Our consistency analysis is then presented in Sect. IV, and a brief outlook follows in Sect. V.

\section{\label{unify}Varying couplings and unification}

The Standard Model (SM) of particle physics is one of the most respected
buildings of modern physics. The desire to push it forward and understand
its correctness at sufficiently high energy scales (the Planck mass, for
instance) raises some open issues. One hint to the fact that the SM might be
incomplete is the running of coupling constants with energy. These
couplings are expected to meet at an energy scale of $M_{GUT} \sim 10^{16}$ GeV.
Considering this fact, it seems reasonable to think of a scenario in which
all interactions stem from only one theory: a Grand Unified Theory (GUT).

In a general GUT, the symmetries of the SM---the gauge group $SU(3) \times
SU(2) \times U(1)$---are unified into a larger symmetry group with only one
coupling parameter at the GUT energy scale $M_{GUT} \sim 10^{16}$ GeV. This
coupling is assumed to be an independent parameter and it is allowed to
vary. Assuming one such GUT, one can relate all couplings of SM
interactions to only one, fundamental coupling. Since the latter can vary,
all other related, lower energy, couplings must also vary and those
variations should be related.

There are many models that attempt to unify the SM interactions. We will be
working with a specific class of models that makes the following assumptions:
the electroweak scale is derived by dimensional transmutation, all Yukawa
couplings vary in the same way and the variation of the couplings is
assumed to be a result of a dilaton-type scalar field. Under these
assumptions it is possible to encapsulate the details of these class of
theories in two parameters: $R$ and $S$ \cite{Campbell,Coc}. They are defined as
proportionality factors between variations of fundamental couplings and
they are highly model dependent:
\begin{equation}
\frac{\Delta \Lambda_{QCD}}{\Lambda_{QCD}} = R \frac{\Delta \alpha}{\alpha}
\end{equation}
\begin{equation}
\frac{\Delta \nu}{\nu} = S \frac{\Delta h}{h}
\end{equation}
where $R \sim 95.7 \frac{\Delta b_3}{5/3 \Delta b_1 + \Delta b_2}$ and $S =
\frac{d \text{ln}\nu}{d \text{ln} h}$, in which $b_i$ is the beta function
coefficient for the $i^{th}$ SM gauge coupling ($i=1,2,3$) at an energy scale smaller than
the unification scale, $\Lambda_{QCD}$ is the QCD mass scale, $\alpha$ is
the fine-structure constant, $\nu$ is the vacuum expectation value of the
Higgs and $h$ stands for any of the Yukawa couplings (we're assuming that
all of them vary in the same way).

Using these two parameters, one can express, in a compact way, the
varying-coupling character of a wide class of theories and, by constraining
them, it is possible to shed some light on the specificities of a putative
GUT. Hence, in what follows, we'll use them to relate the variations of 
interest as \cite{Coc,Luo,Clocks}
\begin{equation}
\frac{\Delta \mu}{\mu} = [0.8 R - 0.3 (1+S)] \frac{\Delta \alpha}{\alpha}
\end{equation}
\begin{equation}
\frac{\Delta g_p}{g_p} = [0.1 R - 0.04(1+S)] \frac{\Delta \alpha}{\alpha}
\end{equation}
\begin{equation}
\frac{\Delta g_n}{g_n} = [0.12 R  - 0.05(1+S)] \frac{\Delta \alpha}{\alpha}\,.
\end{equation}
Note that in this class of models the proton gyromagnetic ratio is less sensitive to the
parameters $R$ and $S$ than the proton-to-electron mass ratio. We will use this
observation in some of our subsequent analysis. We also point out that in this
class of models the behavior of the temperature-redshift relation provides a
further consistency test, as first discussed in \cite{Tofz1,Tofz2}.

\section{\label{qsodata}Current spectroscopic measurements}

We now list the astrophysical measurements that will be used in our analysis. Unless otherwise stated, we will list them in units of parts per million (ppm). This is not meant to be an exhaustive list of all measurements. We typically use only the tightest available measurement for each astrophysical source. A few older measurements along other lines of sight have not been used, on the grounds that they would have no statistical weight in the analysis. Nevertheless, we will include some low-sensitivity but high-redshift measurements, as these are illustrative of the redshift range that may be probed by future facilities. As in \cite{Frigola}, whose list we update here, our two exceptions regarding measurements of the same source are
\begin{itemize}
\item Measurements using different, independent techniques---typically this occurs with measurements of $\mu$ or combined measurements using different molecules, and
\item Measurements obtained with different spectrographs\,.
\end{itemize}
In these cases we do list the various available measurements.

\begin{table}
\begin{tabular}{|c|c|c|c|c|}
\hline
Object & z & $Q_{AB}$  & ${ \Delta Q_{AB}}/{Q_{AB}}$ & Ref. \\ 
\hline\hline
PKS1413$+$135 & 0.247 & ${\alpha^{2\times1.85}g_{p}\mu^{1.85}}$  & $-11.8\pm4.6$ & \protect\cite{Kanekar2} \\
\hline
PKS1413$+$135 & 0.247 & ${\alpha^{2\times1.57}g_{p}\mu^{1.57}}$  & $5.1\pm12.6$ & \protect\cite{Darling} \\
\hline
PKS1413$+$135 & 0.247 & ${\alpha^{2}g_{p}}$  & $-2.0\pm4.4$ & \protect\cite{Murphy} \\
\hline\hline
B0218$+$357 & 0.685 & ${\alpha^{2}g_{p}}$ & $-1.6\pm5.4$ & \protect\cite{Murphy} \\
\hline\hline
J0134$-$0931 & 0.765 & ${\alpha^{2\times1.57}g_{p}\mu^{1.57}}$  &  $-5.2\pm4.3$ & \protect\cite{Kanekar} \\
\hline\hline
J2358$-$1020 & 1.173 & ${\alpha^{2}g_{p}/\mu}$ & $1.8\pm2.7$ & \protect\cite{Rahmani} \\
\hline\hline
J1623$+$0718 & 1.336 & ${\alpha^{2}g_{p}/\mu}$ & $-3.7\pm3.4$ & \protect\cite{Rahmani} \\
\hline\hline
J2340$-$0053 & 1.361 & ${\alpha^{2}g_{p}/\mu}$ & $-1.3\pm2.0$ & \protect\cite{Rahmani} \\
\hline\hline
J0501$-$0159 & 1.561 & ${\alpha^{2}g_{p}/\mu}$ & $3.0\pm3.1$ & \protect\cite{Rahmani} \\
\hline\hline
J1024$+$4709 & 2.285 & ${\alpha^{2}\mu}$ & $100\pm40$ & \protect\cite{Curran} \\
\hline\hline
J2135$-$0102 & 2.326 & ${\alpha^{2}\mu}$ & $-100\pm100$ & \protect\cite{Curran} \\
\hline\hline
J1636$+$6612 & 2.517 & ${\alpha^{2}\mu}$ & $-100\pm120$ & \protect\cite{Curran} \\
\hline\hline
H1413$+$117 & 2.558 & ${\alpha^{2}\mu}$ & $-40\pm80$ & \protect\cite{Curran} \\
\hline\hline
J1401$+$0252 & 2.565 & ${\alpha^{2}\mu}$ & $-140\pm80$ & \protect\cite{Curran} \\
\hline\hline
J0911$+$0551 & 2.796 & ${\alpha^{2}\mu}$ & $ -6.9\pm3.7$ & \protect\cite{Weiss} \\
\hline\hline
J1337$+$3152  & 3.174 & ${\alpha^{2}g_{p}/\mu}$ & $-1.7\pm1.7$ & \protect\cite{Petitjean1} \\
\hline\hline
APM0828$+$5255 & 3.913 & ${\alpha^{2}\mu}$ & $-360\pm90$ & \protect\cite{Curran} \\
\hline\hline
MM1842$+$5938 & 3.930 & ${\alpha^{2}\mu}$ & $-180\pm40$ & \protect\cite{Curran} \\
\hline\hline
PSS2322$+$1944 & 4.112 & ${\alpha^{2}\mu}$ & $170\pm130$ & \protect\cite{Curran} \\
\hline\hline
BR1202$-$0725 & 4.695 & ${\alpha^{2}\mu}$ & $50\pm150$ & \protect\cite{Lentati} \\
\hline\hline
J0918$+$5142 & 5.245 & ${\alpha^{2}\mu}$ & $-1.7\pm8.5$ & \protect\cite{Levshakov} \\
\hline\hline
J1148$+$5251 & 6.420 & ${\alpha^{2}\mu}$ & $330\pm250$ & \protect\cite{Lentati} \\
\hline\hline\hline
\end{tabular}
\caption{\label{table1}Available measurements of several combinations of the dimensionless couplings $\alpha$, $\mu$ and $g_p$. Listed are, respectively, the object along each line of sight, the redshift of the measurement, the dimensionless parameter being constrained, the measurement itself (in parts per million), and its original reference.}
\end{table}

Table \ref{table1} contains current joint measurements of several couplings. Compared to our earlier work we have added the measurements of \cite{Curran}, which although not particularly sensitive complement the other measurements in terms of redshift coverage. (We thank Hugo Messias for bringing this reference to our attention.) Note that for the radio source PKS1413$+$135 the three available measurements are sufficient to yield individual constraints on the variations of the three quantities at redshift $z=0.247$. This analysis was done in \cite{PKS}, yielding a null result at the two sigma confidence level.

\begin{table}
\begin{center}
\begin{tabular}{|c|c|c|c|c|}
\hline
 Object & z & ${ \Delta\alpha}/{\alpha}$ (ppm) & Spectrograph & Ref. \\
\hline\hline
3 sources & 1.08 & $4.3\pm3.4$ & HIRES & \protect\cite{Songaila} \\
\hline
HS1549$+$1919 & 1.14 & $-7.5\pm5.5$ & UVES/HIRES/HDS & \protect\cite{LP3} \\
\hline
HE0515$-$4414 & 1.15 & $-0.1\pm1.8$ & UVES & \protect\cite{alphaMolaro} \\
\hline
HE0515$-$4414 & 1.15 & $0.5\pm2.4$ & HARPS/UVES & \protect\cite{alphaChand} \\
\hline
HS1549$+$1919 & 1.34 & $-0.7\pm6.6$ & UVES/HIRES/HDS & \protect\cite{LP3} \\
\hline
HE0001$-$2340 & 1.58 & $-1.5\pm2.6$ &  UVES & \protect\cite{alphaAgafonova}\\
\hline
HE1104$-$1805A & 1.66 & $-4.7\pm5.3$ & HIRES & \protect\cite{Songaila} \\
\hline
HE2217$-$2818 & 1.69 & $1.3\pm2.6$ &  UVES & \protect\cite{LP1}\\
\hline
HS1946$+$7658 & 1.74 & $-7.9\pm6.2$ & HIRES & \protect\cite{Songaila} \\
\hline
HS1549$+$1919 & 1.80 & $-6.4\pm7.2$ & UVES/HIRES/HDS & \protect\cite{LP3} \\
\hline
Q1101$-$264 & 1.84 & $5.7\pm2.7$ &  UVES & \protect\cite{alphaMolaro}\\
\hline
\end{tabular}
\caption{\label{table2}Available specific measurements of $\alpha$. Listed are, respectively, the object along each line of sight, the redshift of the measurement, the measurement itself (in parts per million), the spectrograph, and the original reference. The recent UVES Large Program measurements are \cite{LP1,LP3}. The first measurement is the weighted average from 8 absorbers in the redshift range $0.73<z<1.53$ along the lines of sight of HE1104-1805A, HS1700+6416 and HS1946+7658, reported in \cite{Songaila} without the values for individual systems. The UVES, HARPS, HIRES and HDS spectrographs are respectively in the VLT, ESO 3.6m, Keck and Subaru telescopes.}
\end{center}
\end{table}

Table \ref{table2} contains individual $\alpha$ measurements. Conservatively we only list measurements where data was acquired specifically for this purpose---but these are by now, in most cases, the ones with the smallest uncertainties. Compared to our previous analysis we have added the most recent Large Program result \cite{LP3} as well as the measurements of \cite{Songaila}. We note that the weighted mean of the measurements on the table is
\begin{equation}\label{prioralpha}
\left(\frac{\Delta\alpha}{\alpha}\right)_{wm}=0.37\pm0.94\,,
\end{equation}
and thus, for the first time, the one-sigma statistical uncertainty is below ppm (although the key issue is that of possible systematics).

\begin{table}
\begin{tabular}{|c|c|c|c|c|}
\hline
 Object & z & ${\Delta\mu}/{\mu}$ & Method & Ref. \\ 
\hline\hline
B0218$+$357 & 0.685 & $0.74\pm0.89$ & $NH_3$/$HCO^+$/$HCN$ & \protect\cite{Murphy2} \\
\hline
B0218$+$357 & 0.685 & $-0.35\pm0.12$ & $NH_3$/$CS$/$H_2CO$ & \protect\cite{Kanekar3} \\
\hline\hline
PKS1830$-$211 & 0.886 & $0.08\pm0.47$ &  $NH_3$/$HC_3N$ & \protect\cite{Henkel}\\
\hline
PKS1830$-$211 & 0.886 & $-1.2\pm4.5$ &  $CH_3NH_2$ & \protect\cite{Ilyushin}\\
\hline
PKS1830$-$211 & 0.886 & $-2.04\pm0.74$ & $NH_3$ & \protect\cite{Muller}\\
\hline
PKS1830$-$211 & 0.886 & $-0.10\pm0.13$ &  $CH_3OH$ & \protect\cite{Bagdonaite2}\\
\hline\hline
J2123$-$005 & 2.059 & $8.5\pm4.2$ & $H_2$/$HD$ (VLT)& \protect\cite{vanWeerd} \\
\hline
J2123$-$005 & 2.059 & $5.6\pm6.2$ & $H_2$/$HD$ (Keck)& \protect\cite{Malec} \\
\hline\hline
HE0027$-$1836 & 2.402 & $-7.6\pm10.2$ & $H_2$ & \protect\cite{LP2} \\
\hline\hline
Q2348$-$011 & 2.426 & $-6.8\pm27.8$ & $H_2$ & \protect\cite{Bagdonaite} \\
\hline\hline
Q0405$-$443 & 2.597 & $10.1\pm6.2$ & $H_2$ & \protect\cite{King} \\
\hline\hline
J0643$-$504 & 2.659 & $7.4\pm6.7$ & $H_2$ & \protect\cite{Albornoz} \\
\hline\hline
Q0528$-$250 & 2.811 & $0.3\pm3.7$ & $H_2$/$HD$ & \protect\cite{King2} \\
\hline\hline
Q0347$-$383 & 3.025 & $2.1\pm6.0$ & $H_2$ & \protect\cite{Wendt} \\
\hline\hline
J1443$+$2724 & 4.224 & $-9.5\pm7.6$ & $H_2$ & \protect\cite{Bagdonaite4} \\
\hline\hline\hline
\end{tabular}
\caption{\label{table3}Available measurements of $\mu$. Listed are, respectively, the object along each line of sight, the redshift of the measurement, the measurement itself, the molecule(s) used, and the original reference. The recent LP measurement is \cite{LP2}.}
\end{table}

Table \ref{table3} contains individual $\mu$ measurements. Note that several different molecules can be used, and in the case of the gravitational lens PKS1830$-$211 several independent measurements exist. Currently ammonia is the most common at low redshift, though others such as methanol, peroxide, hydronium and methanetiol have a greater potential in the context of facilities like ALMA \cite{Molecules}. At higher redshifts molecular hydrogen is the most common.

The tightest available constraint was obtained in PKS1830$-$211, from observations of methanol transitions \cite{Bagdonaite2}. A recent analysis leads to a nominally tighter bound \cite{BagdonaiteNew2}, but the authors suggest caution in using that result due to possible systematics; conservatively, we have therefore kept the earlier measurement. As for measurements along the line of sight towards J0643$-$504, we have similarly listed the more conservative one, from \cite{Albornoz} on the table; for comparison \cite{BagdonaiteNew} finds $\Delta\mu/\mu=12.7\pm6.2$ ppm.

The main update to these measurements is therefore the arrival of the first direct constraint on $\mu$ beyond $z=4$ \cite{Bagdonaite4}. Calculating the weighted mean of the low and high-redshift samples ($z<1$ and $z>2$, respectively, we find
\begin{equation}\label{priormulo}
\left(\frac{\Delta\mu}{\mu}\right)_{Low,wm}=-0.24\pm0.09\,
\end{equation}
\begin{equation}\label{priormuhi}
\left(\frac{\Delta\mu}{\mu}\right)_{High,wm}=3.4\pm2.0\,,
\end{equation}
in both cases this is a very mild evidence for a variation, though with different signs at high and low redshifts,

\section{\label{newres}Consistency analysis}

\begin{figure*}
\begin{center}
\includegraphics[width=3in]{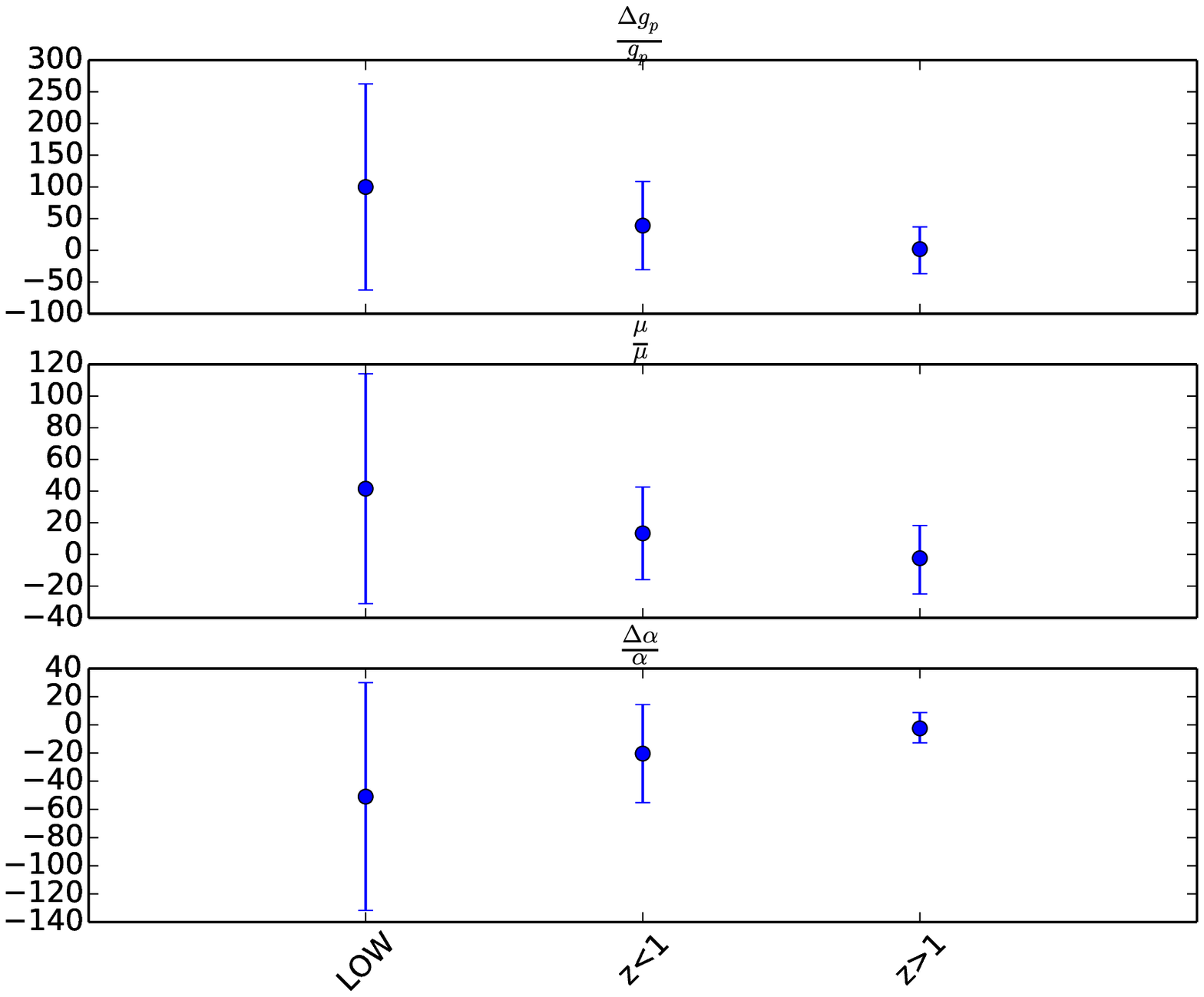}
\includegraphics[width=3in]{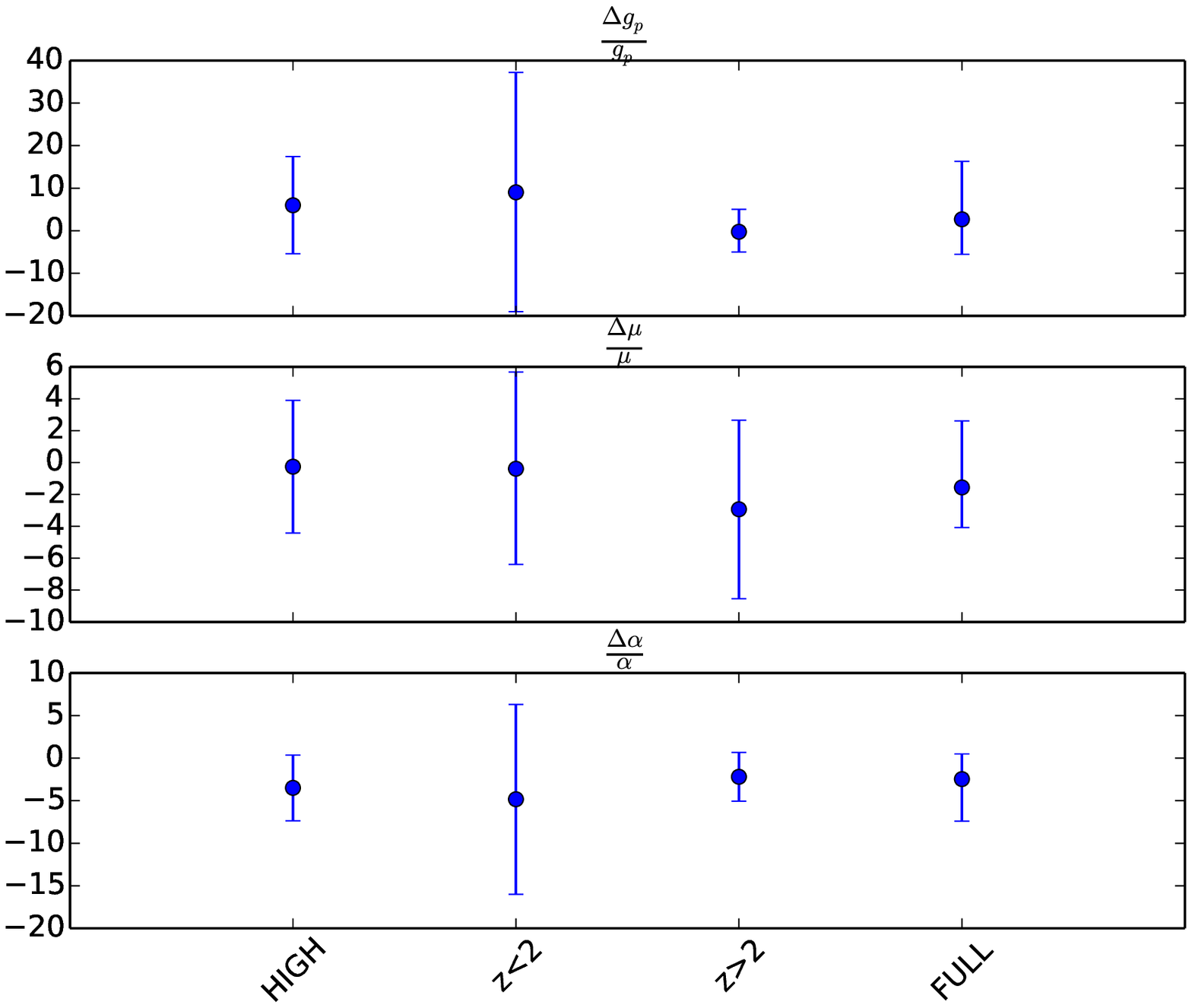}
\end{center}
\caption{\label{fig1}Best-fit parameters for the data in \protect\ref{table1}: One-dimensional confidence intervals for the relative variations of the three couplings (in ppm), for the full sample and the subsamples discussed in the text.} Note that the vertical scale is different in both sets of panels.
\end{figure*}
\begin{figure}
\begin{center}
\includegraphics[width=3in]{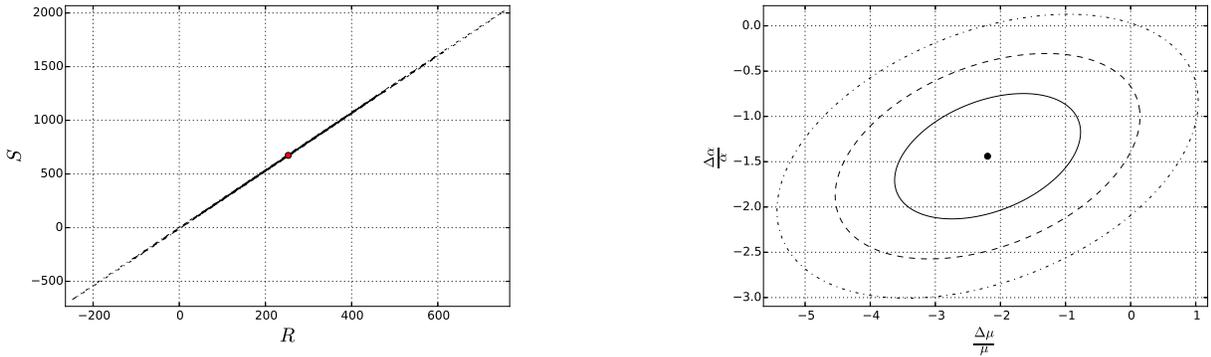}
\end{center}
\caption{\label{fig2}Best-fit parameters for the data in \protect\ref{table1}: Two-dimensional one, two and three sigma confidence intervals in the R-S plane.}
\end{figure}
\begin{figure}
\begin{center}
\includegraphics[width=3in]{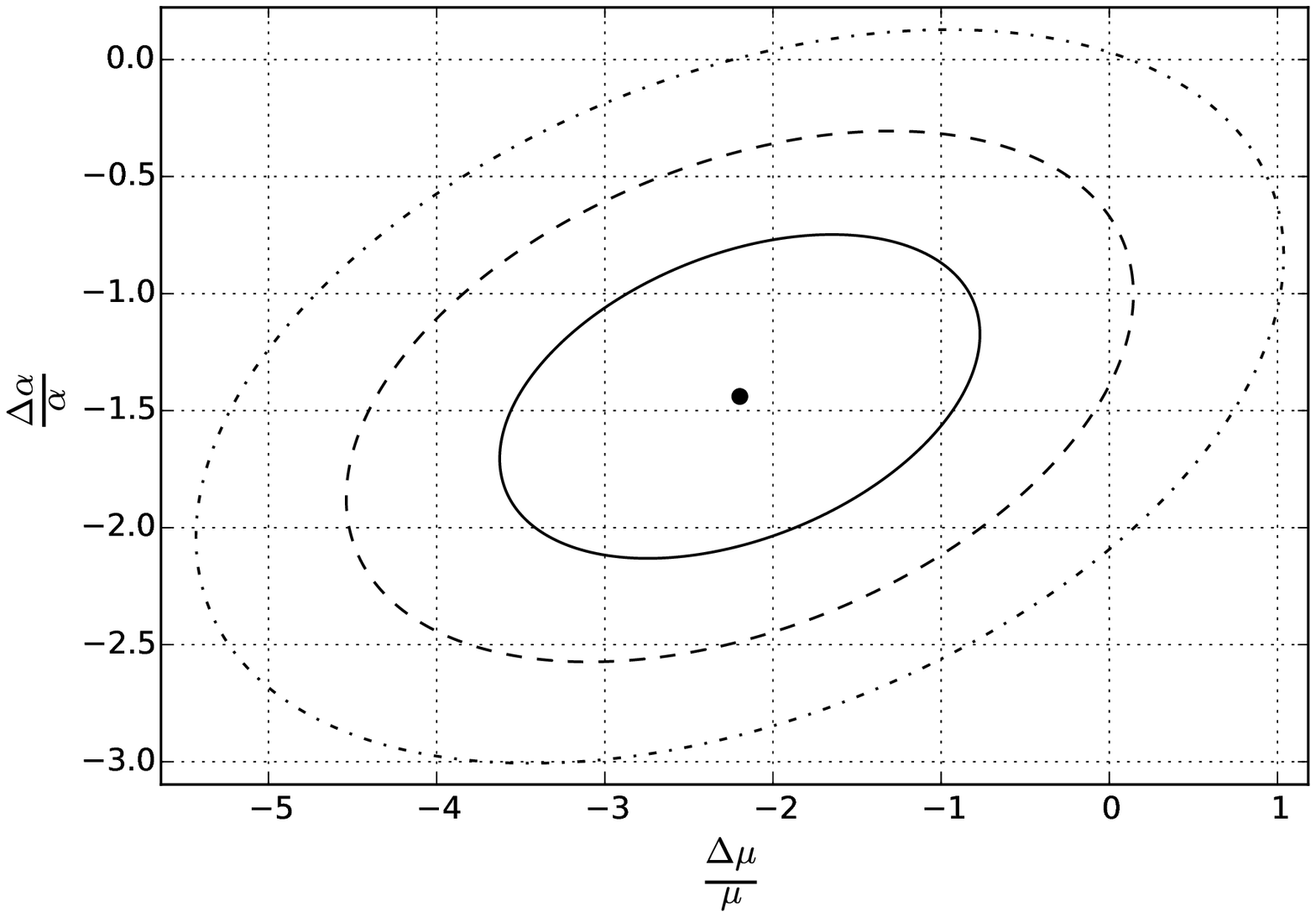}
\includegraphics[width=3in]{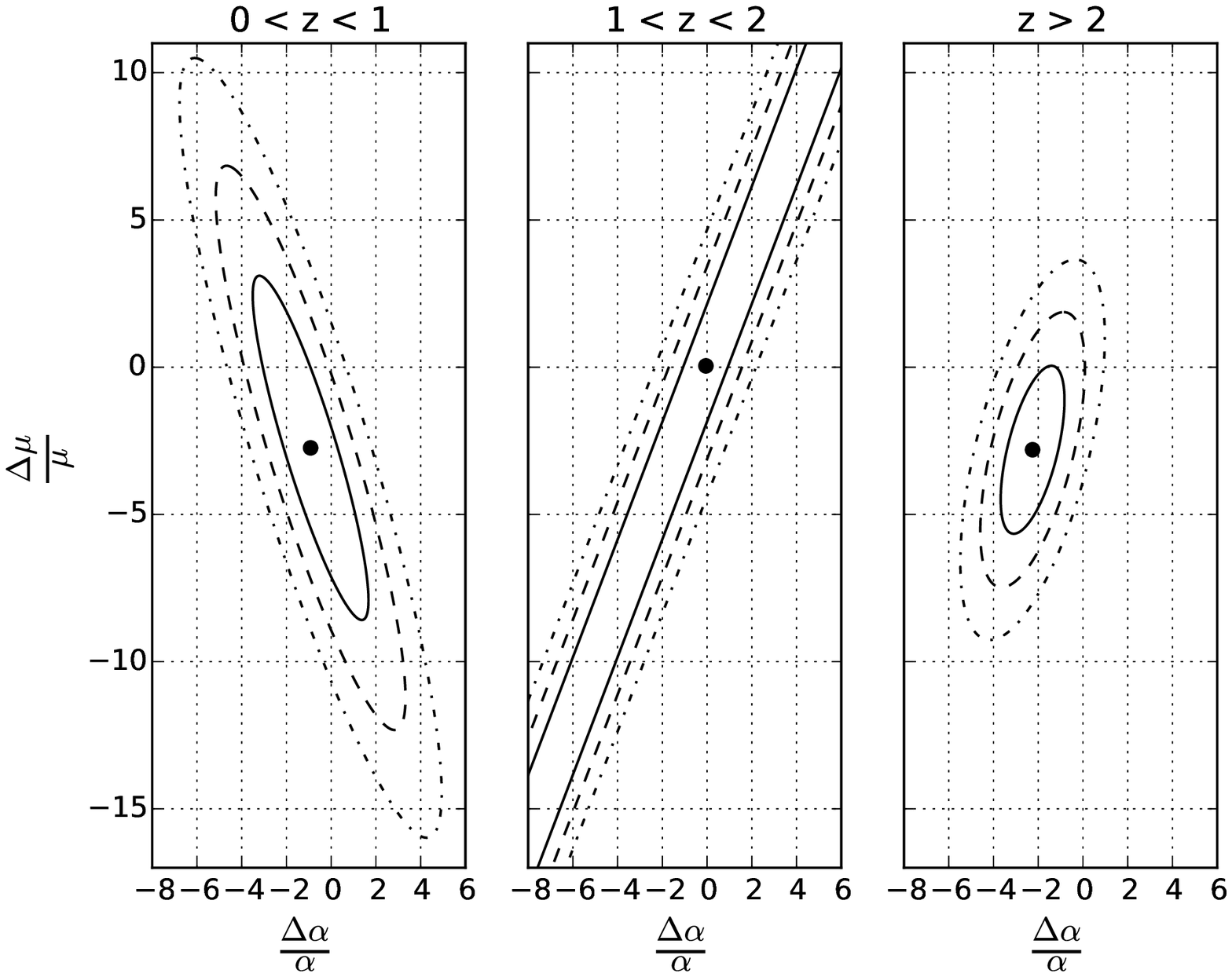}
\end{center}
\caption{\label{fig4a2}Two-dimensional likelihood in the $\alpha-\mu$ plane for the data in \protect\ref{table1}, assuming no variations of $g_p$, for the full table (top panel) and dividing the data into three redshift bins (bottom panel). In both cases the one, two and three sigma confidence intervals are plotted.}
\end{figure}

We start by determining the individual values of the three couplings that provide the best fit to the data in Table \ref{table1}. The results are summarized in Fig. \ref{fig1}. We obtained the following one-sigma constraints for the full table
\begin{equation}
\frac{\Delta\alpha}{\alpha}=-2.5^{+3.2}_{-5.2}\,
\end{equation}
\begin{equation}
\frac{\Delta\mu}{\mu}=-1.6^{+4.2}_{-2.7}\,
\end{equation}
\begin{equation}
\frac{\Delta g_p}{g_p}=2.7^{+13.8}_{-8.4}\,
\end{equation}
thus consistent with no variations. In addition to the full sample the figure also shows the results of the analysis for a few subsamples
\begin{itemize}
\item $z<0.25$ (denoted {\bf Low} in the plot, and containing only the measurements along the line-of-sight to PKS1413$+$135 discussed in \cite{PKS}) versus $z>0.25$ (denoted {\bf High} in the plot)
\item $z<1$ versus $z>1$
\item $z<2$ versus $z>2$
\end{itemize}
We note that the likelihood profiles are usually non-Gaussian, both here and in the analyses in the rest of this section. This explains why some of the confidence intervals are asymmetric.

Fig. \ref{fig2} shows the two-dimensional confidence levels in the plane of the unification parameters R and S discussed in Sect. II. The one dimensional marginalized likelihoods for each of them are
\begin{equation}
R=253\pm221\,,\quad S=673\pm594\,.
\end{equation}
These are fully consistent with the naive expectations suggesting typical values around $R\sim30$ and $S\sim160$ \cite{Coc} as well as with the results of our earlier work \cite{Frigola}.

\begin{table}
\begin{center}
\begin{tabular}{|c|c|c|}
\hline
 Sample & ${ \Delta\alpha}/{\alpha}$ & ${ \Delta\mu}/{\mu}$ \\
\hline\hline
$z<1$ &  $-0.9\pm4.2$ & $-2.7\pm9.6$ \\
\hline
$1<z<2$ & Unconstrained & Unconstrained \\
\hline
$z>2$ & $-2.3\pm2.3$ & $-2.8\pm4.7$ \\
\hline
Full & $-1.4\pm1.1$ & $-2.2\pm2.3$\\
\hline
\end{tabular}
\caption{\label{tablefix}One-dimensional marginalized two-sigma constraints for $\alpha$ and $\mu$, for the data in \protect\ref{table1}, assuming no variations of $g_p$, for the full sample and various redshift bins. All constraints are in parts per million.}
\end{center}
\end{table}

Since in this class of models the variations of $g_p$ are, for typical values of the parameters $R$ and $S$, smaller than those of $\mu$, it's interesting to repeat some of the above analysis on the assumption that $g_p$ does not vary. These results are shown in Fig. \ref{fig4a2}, and the 1D marginalized constraints for the full sample and various subsamples are shown in Table \ref{tablefix}. 

Note that in this case there are again deviations (at up to the two sigma level) from the null result, which wasn't the case in the full analysis where $g_p$ was also allowed to vary. This change stems from the reduction in uncertainties afforded by the assumption of a fixed $g_p$ (the shift in best-fit values is comparatively small), and is therefore not statistically significant. In the redshift range $1<z<2$ the individual values of both parameters are unconstrained by this dataset: only the combination $\alpha^2/\mu$ is constrained, but this constraint does play a role in the full likelihood for this case. Finally, it's also interesting to note that the degeneracy directions are different at high and low redshifts.

\begin{figure}
\begin{center}
\includegraphics[width=3in]{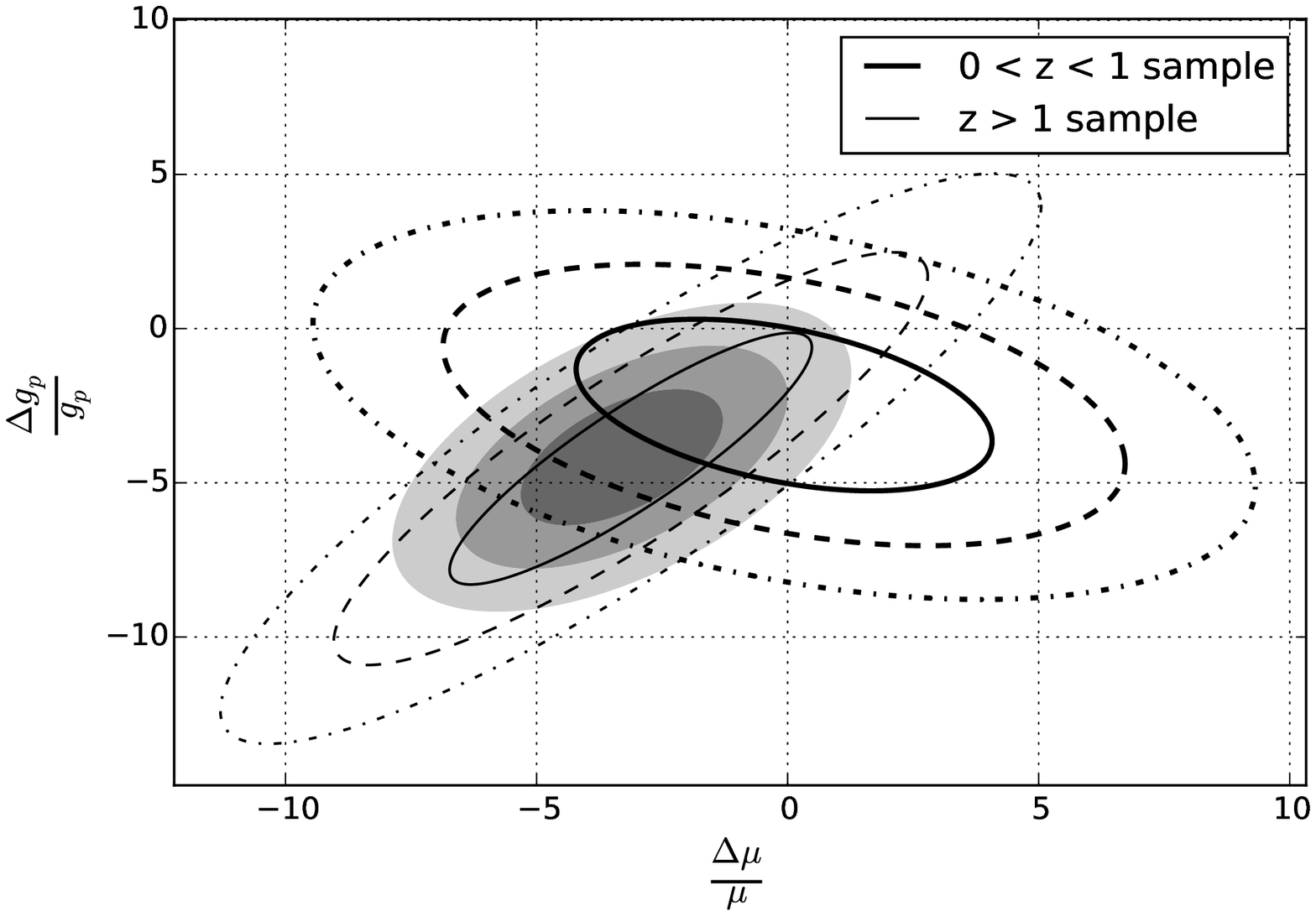}
\includegraphics[width=3in]{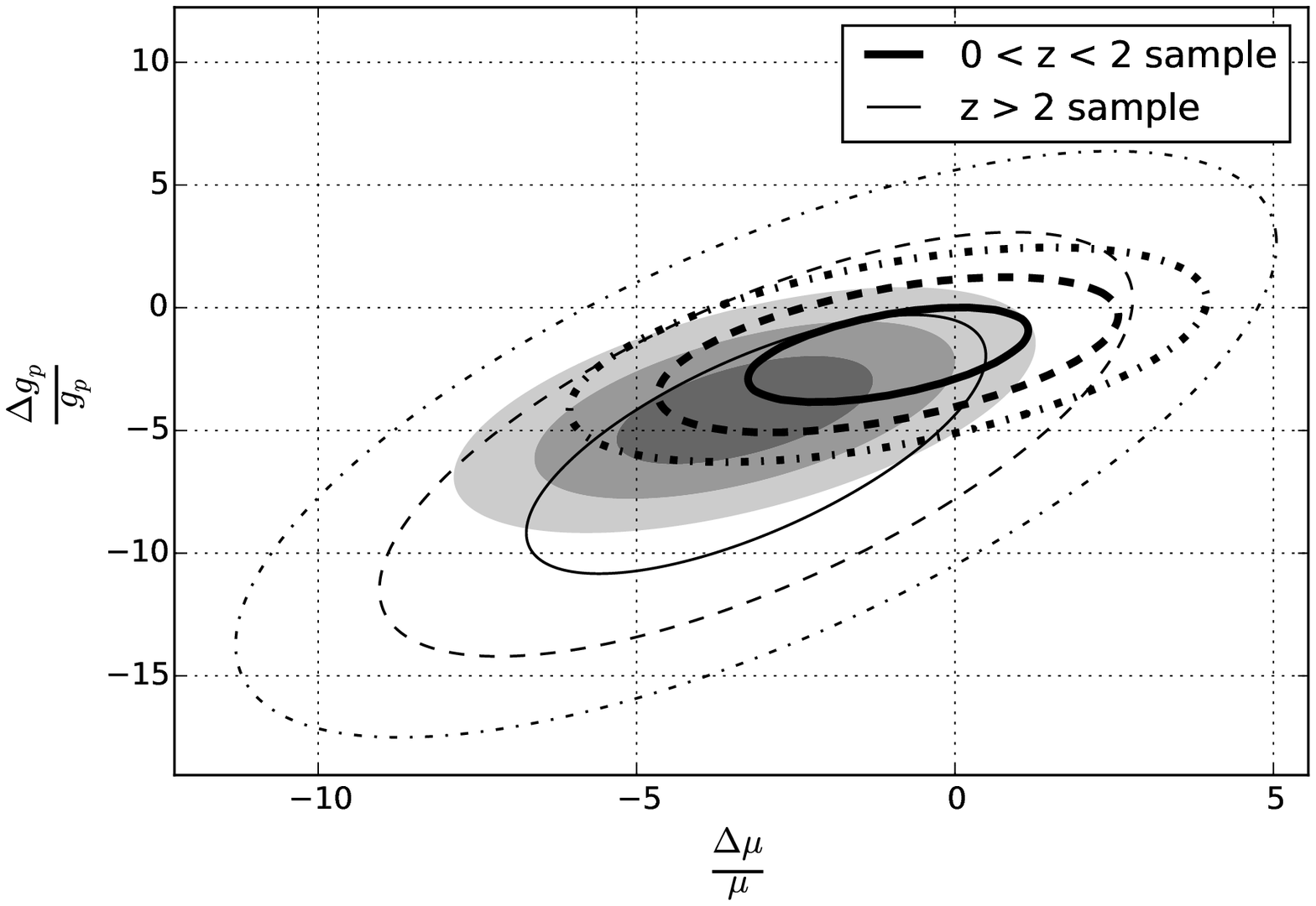}
\end{center}
\caption{\label{old5}Two-dimensional likelihood (one, two and three sigma confidence levels) in the $\mu-g_p$ plane for the data in Table \protect\ref{table1} assuming the weighed mean value of Table \protect\ref{table2} for $\alpha$ (given by Eq.\protect\ref{prioralpha}).In both panels the two sets of black contours show the likelihood for low and high redshift subsamples, and the shaded regions show the likelihood for the full sample. In the top panel the borderline between the two subsamples is at $z=1$, while in the bottom one it's at $z=2$.}
\end{figure}

We can also reanalyze the data of Table \ref{table1} using the weighed mean value of Table \ref{table2} for $\alpha$ (ie, Eq.\protect\ref{prioralpha}) as a prior. The results of this analysis are summarized in Fig. \ref{old5} and Table \ref{tablepria}, both for the full sample and for high/low redshift subsamples. There is a mild (two-sigma) preference for negative variations of both quantities, which mostly stems from the higher redshift data. As before, note that the degeneracy directions change with redshift; this partly explains why the preference for negative variations is much stronger for the full sample than the subsamples. On the other hand, it could also be an indication that assuming a constant value for the parameters in this entire redshift range may be too simplistic.

\begin{table}
\begin{center}
\begin{tabular}{|c|c|c|}
\hline
 Sample & ${ \Delta\mu}/{\mu}$ & ${ \Delta g_p}/{g_p}$ \\
\hline\hline
$0<z<1$ &  $-0.1\pm6.8$ & $-2.5\pm4.5$ \\
\hline
$z>1$ & $-3.1\pm5.9$ & $-4.2\pm6.7$ \\
\hline
$0<z<2$ & $-1.1\pm3.6$ & $-1.9\pm3.2$ \\
\hline
$z>2$ & $-3.1\pm5.9$ & $-5.6\pm8.6$ \\
\hline
Full & $-3.3\pm3.3$ & $-4.2\pm3.6$\\
\hline
\end{tabular}
\caption{\label{tablepria}One-dimensional marginalized two-sigma constraints for $\mu$ and $g_p$, for the data in Table \protect\ref{table1}, assuming Eq. \protect\ref{prioralpha} as prior on $\alpha$, for the full sample and various redshift bins. All constraints are in parts per million.}
\end{center}
\end{table}

\begin{figure}
\begin{center}
\includegraphics[width=3in]{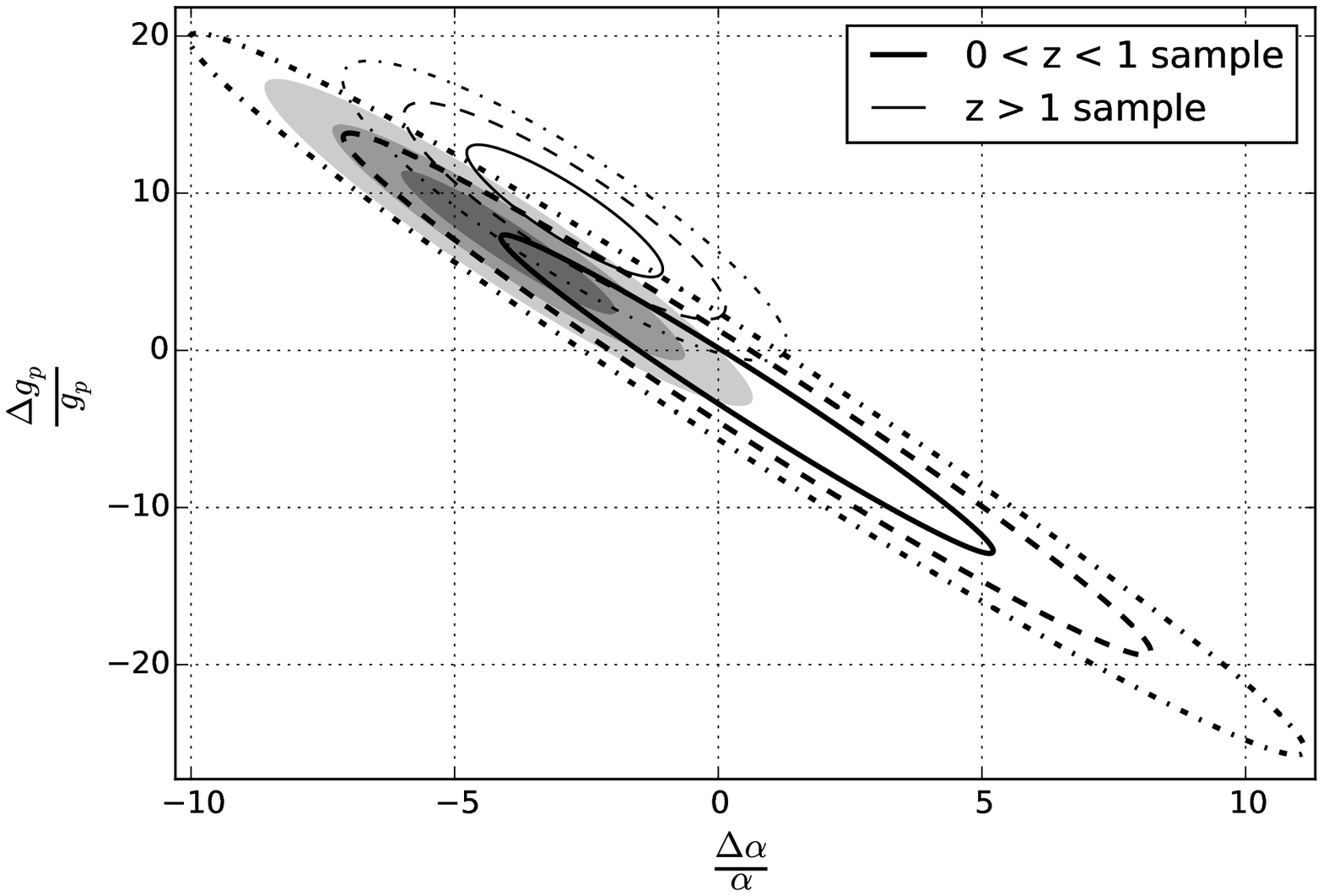}
\includegraphics[width=3in]{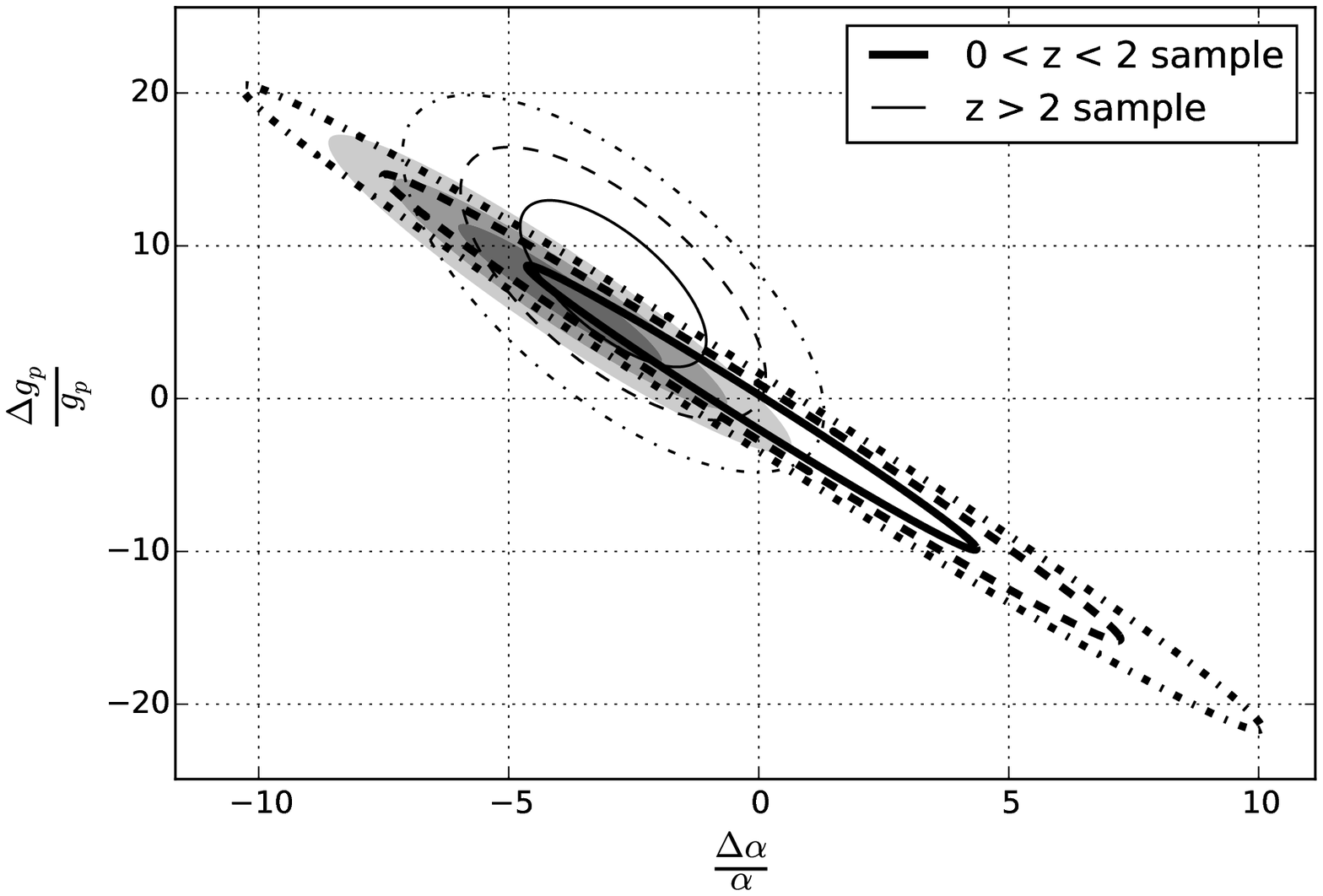}
\end{center}
\caption{\label{old7}Two-dimensional likelihood (one, two and three sigma confidence levels) in the $\alpha-g_p$ plane for the data in Table \protect\ref{table1} assuming the weighed mean values of Table \protect\ref{table3} at low and high redshifts for $\mu$ (cf. Eqs. \protect\ref{priormulo}-\ref{priormuhi}).In both panels the two sets of black contours show the likelihood for low and high redshift subsamples, and the shaded regions show the likelihood for the full sample. In the top panel the borderline between the two subsamples is at $z=1$, while in the bottom one it's at $z=2$.}
\end{figure}

The complementary exercise can also be done: we can analyze the data of Table \ref{table1} using the weighed mean values of Table \ref{table3} for $\mu$ at low and high redshifts (ie, Eqs.\ref{priormulo}-\ref{priormuhi}) as priors. The corresponding results are summarized in Fig. \ref{old7} and Table \ref{tableprim}. We again considered both the full sample and high and low redshift subsamples. There is again a mild (two-sigma) preference for variations of both quantities, which again mostly stems from the higher redshift data. In this case the results are not as clear which is at least in part the result of a stronger degeneracy than in the previous case. One effect of the $\mu$ priors (with their slight preference for negative variations) is to lead to a preference for positive values of $g_p$ (while a negative value is preferred for $\alpha$).

\begin{table}
\begin{center}
\begin{tabular}{|c|c|c|}
\hline
 Sample & ${ \Delta\alpha}/{\alpha}$ & ${ \Delta g_p}/{g_p}$ \\
\hline\hline
$0<z<1$ &  $0.5\pm7.6$ & $-2.8\pm16.6$ \\
\hline
$z>1$ & $-2.9\pm3.0$ & $8.9\pm6.9$ \\
\hline
$0<z<2$ & $-0.1\pm7.4$ & $-0.6\pm15.2$ \\
\hline
$z>2$ & $-2.9\pm3.0$ & $7.5\pm8.9$ \\
\hline
Full & $-4.0\pm3.3$ & $6.9\pm7.5$\\
\hline
\end{tabular}
\caption{\label{tableprim}One-dimensional marginalized two-sigma constraints for $\alpha$ and $g_p$, for the data in Table \protect\ref{table1}, assuming Eqs. \protect\ref{priormulo}-\protect\ref{priormuhi} as low and high-redshift priors on $\mu$, for the full sample and various redshift bins. All constraints are in parts per million.}
\end{center}
\end{table}

As an additional check, if we assume that $g_p$ does not vary and use the weighed mean of Table \ref{table2} for $\frac{\Delta \alpha}{\alpha}$ as a prior for Table \ref{table1} we now find, at the two-sigma confidence level,
\begin{equation}
\frac{\Delta\mu}{\mu}=-2.9\pm2.2\;
\end{equation}
compared to the more general analysis summarized in Table \ref{tablepria} the central value is in good agreement, while the uncertainty is naturally reduced. Conversely, if we assume that $g_p$ is fixed and use the weighted mean value of Table \ref{table3} for $\frac{\Delta \mu}{\mu}$ we find, still at the two-sigma level
\begin{equation}
\frac{\Delta\alpha}{\alpha}=-1.1\pm0.8\,;
\end{equation}
again, this is consistent with the results of Table \ref{tableprim}, though with a significantly reduced uncertainty. Finally if we simultaneously use the weighted mean averages of Tables \ref{table2} and \ref{table3} as priors for $\frac{\Delta \alpha}{\alpha}$ and $\frac{\Delta \mu}{\mu}$ and use Table \ref{table1} to infer $g_p$, we obtain at the two-sigma confidence level
\begin{equation}
\frac{\Delta g_p}{g_p}=-2.1\pm2.8\,.
\end{equation}

\section{\label{concl}Conclusions}

In this work we have fully updated an earlier analysis \cite{Frigola} of the consistency of the various currently available astrophysical tests of the stability of fundamental couplings. Various combinations of $\alpha$, $\mu$ and $g_p$ can be measured in the radio band, while $\alpha$ and $\mu$ can also be individually measured, mostly through observations in the optical/ultraviolet region. Taken together, they span very broad redshift range---approximately from $z=0.2$ to $z=6.4$---although the sensitivity of current measurements is also quite heterogeneous: from about 0.1 ppm to more than 100 ppm.

Our results show that a joint analysis of all the combined measurements is consistent with no variations. However, joint analyses of the radio and optical/UV data tend to lead to mild evidence (specifically, at two-sigma confidence level) for variations at the few ppm level. This is especially the case at redshifts $z>1$, corresponding to measurements in the matter era.

It is noteworthy that the sensitivity with which each parameter can be constrained is redshift dependent. The best direct individual constraints on $\alpha$ (at the few ppm level) are currently in the redshift range $1<z<2$, while those for $\mu$ are either at $z>2$ (from molecular Hydrogen measurements) or at $z<1$ (from measurements using other molecules, such as ammonia or methanol). Similarly, different products of $\alpha$, $\mu$ and $g_p$ can be constrained at different redshift. This is relevant for optimizing observational strategies and selecting targets for future facilities.

Admittedly the evidence for possible variations thus inferred is not strong, and the most likely explanation for them is that systematic errors have not been fully accounted for. In the case of the optical measurements, possible sources for these have recently been studied in some detail \cite{LP1,LP2,LP3,Syst}. In any case, clarifying these discrepancies is essential. Finding lines of sight where these measurements can be carried out both in the optical and in the radio bands would provide an ideal test of possible systematics, but the number of such targets is likely to be small.

A more immediate goal will be to extend the range of redshifts where sensitive measurements of $\alpha$ and $\mu$ can be made. The imminent arrival of observational facilities such as ALMA and ESPRESSO will make this possible. This will signal the start of a new era of precision consistency tests of the current cosmological paradigm, which will continue with the E-ELT. A roadmap for these tests can be found in \cite{grg}.

\begin{acknowledgments}
This work was done in the context of the project PTDC/FIS/111725/2009 from FCT (Portugal). C.J.M. is also supported by an FCT Research Professorship, contract reference IF/00064/2012, funded by FCT/MCTES (Portugal) and POPH/FSE (EC).
\end{acknowledgments}

\bibliography{qso}

\end{document}